\documentclass[fleqn,usenatbib]{mnras}

\usepackage{newtxtext,newtxmath}
\usepackage{hyperref}
\usepackage{academicons}
\usepackage{xcolor}

\usepackage[T1]{fontenc}
\usepackage{enumerate}
\DeclareRobustCommand{\VAN}[3]{#2}
\let\VANthebibliography\thebibliography
\def\thebibliography{\DeclareRobustCommand{\VAN}[3]{##3}\VANthebibliography}
\newcommand{\orcid}[1]{\href{https://orcid.org/#1}{\textcolor[HTML]{A6CE39}{\aiOrcid}}}

\usepackage{graphicx}	
\usepackage{amsmath}	
\usepackage{comment}

\newcommand{\swiftj}{SWIFT J1749.4$-$2807}
\newcommand{\nicer}{NICER}
\newcommand{\xmm}{XMM-Newton}
\newcommand{\rxte}{RXTE}
\newcommand{\chandra}{Chandra}
\newcommand{\swift}{Swift}
\newcommand{\inte}{INTEGRAL}

\title[\swiftj{}: long-term evolution]{On the peculiar long-term orbital evolution of the eclipsing accreting millisecond X-ray pulsar \swiftj{}}

\author[A. Sanna et al.]{
A.~Sanna$^{1}$\thanks{E-mail: andrea.sanna@dsf.unica.it} \href{https://orcid.org/0000-0002-0118-2649}{\includegraphics[scale=0.08]{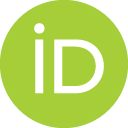}}, L.~ Burderi$^{1}$\href{https://orcid.org/0000-0001-5458-891X}{\includegraphics[scale=0.08]{ORCIDid}}, T.~Di Salvo$^{2}$\href{https://orcid.org/0000-0002-3220-6375}{\includegraphics[scale=0.08]{ORCIDid}}, A.~ Riggio$^{1,3}$\href{https://orcid.org/0000-0002-6145-9224}{\includegraphics[scale=0.08]{ORCIDid}}, D.~ Altamirano$^{4}$\href{https://orcid.org/0000-0002-3422-0074}{\includegraphics[scale=0.08]{ORCIDid}}, A.~ Marino$^{2,13,14}$\href{https://orcid.org/0000-0001-5674-4664}{\includegraphics[scale=0.08]{ORCIDid}}, \newauthor
P.~Bult$^{5,6}$, T.~E.~Strohmayer$^{6}$ \href{https://orcid.org/0000-0001-7681-5845}{\includegraphics[scale=0.08]{ORCIDid}},
S.~Guillot$^{7,8}$\href{https://orcid.org/0000-0002-6449-106X}{\includegraphics[scale=0.08]{ORCIDid}}, C.~Malacaria$^{9}$\href{https://orcid.org/0000-0002-0380-0041}{\includegraphics[scale=0.08]{ORCIDid}}, M.~Ng$^{10}$\href{https://orcid.org/0000-0002-0940-6563}{\includegraphics[scale=0.08]{ORCIDid}}, G.~Mancuso$^{11,12}$\href{https://orcid.org/0000-0001-9822-6937}{\includegraphics[scale=0.08]{ORCIDid}},\newauthor
 S.~M.~Mazzola$^{1}$ \href{https://orcid.org/0000-0001-9230-309X}{\includegraphics[scale=0.08]{ORCIDid}}, A.~C.~Albayati$^{4}$, R. Iaria$^{2}$ \href{https://orcid.org/0000-0003-2882-0927}{\includegraphics[scale=0.08]{ORCIDid}}, A.~Manca$^{1}$, N.~Deiosso$^{1}$, C.~Cabras$^{1}$, A.~Anitra$^{2}$ \href{https://orcid.org/0000-0002-2701-2998}{\includegraphics[scale=0.08]{ORCIDid}} 
\\
$^{1}$Dipartimento di Fisica, Universit\`a degli Studi di Cagliari, SP Monserrato-Sestu km 0.7, 09042 Monserrato, Italy\\
$^{2}$Universit\`a degli Studi di Palermo, Dipartimento di Fisica e Chimica, via Archirafi 36, 90123 Palermo, Italy\\
$^{3}$INAF/IASF Palermo, via Ugo La Malfa 153, I-90146 - Palermo, Italy\\
$^{4}$School of Physics and Astronomy, University of Southampton, Southampton, Hampshire SO17 1BJ, UK\\
$^{5}$Department of Astronomy, University of Maryland, College Park, MD 20742, USA\\
$^{6}$Astrophysics Science Division and Joint Space-Science Institute, NASA Goddard Space Flight Center, Greenbelt, MD 20771, USA\\
$^{7}$IRAP, CNRS, 9 avenue du Colonel Roche, BP 44346, F-31028 Toulouse Cedex 4, France\\
$^{8}$Universit\'e de Toulouse, CNES, UPS-OMP, F-31028 Toulouse, France\\
$^{9}$Universities Space Research Association, Science and Technology Institute, 320 Sparkman Drive, Huntsville, AL 35805, USA\\
$^{10}$MIT Kavli Institute for Astrophysics and Space Research, Massachusetts Institute of Technology, Cambridge, MA 02139, USA\\
$^{11}$Instituto Argentino de Radioastronom\'ia (CCT-La Plata, CONICET; CICPBA), C.C. No. 5, Villa Elisa 1894\\
$^{12}$Argentina; Facultad de Ciencias Astron\'omicas y Geofísicas, Universidad Nacional de La Plata, Paseo del Bosque s/n, La Plata 1900 , Argentina\\
$^{13}$ Institute of Space Sciences (ICE, CSIC), Campus UAB, Carrer de Can Magrans s/n, E-08193 Barcelona, Spain \\
$^{14}$ Institut d'Estudis Espacials de Catalunya (IEEC), E-08034 Barcelona, Spain \\
}

\date{Accepted 2022 June 7. Received 2022 May 10; in original form 2022 February 2}

\pubyear{2022}

\begin{document}
\label{firstpage}
\pagerange{\pageref{firstpage}--\pageref{lastpage}}
\maketitle

\begin{abstract}

We present the pulsar timing analysis of the accreting millisecond X-ray pulsar \swiftj{} monitored by \nicer{} and \xmm{} during its latest outburst after almost eleven years of quiescence. From the coherent timing analysis of the pulse profiles, we updated the orbital ephemerides of the system. Large phase jumps of the fundamental frequency phase of the signal are visible during the outburst, consistent with what was observed during the previous outburst. Moreover, we report on the marginally significant evidence for non-zero eccentricity ($e\simeq 4\times 10^{-5}$) obtained independently from the analysis of both the 2021 and 2010 outbursts and we discuss possible compatible scenarios.
Long-term orbital evolution of \swiftj{} suggests a fast expansion of both the NS projected semi-major axis $(x)$, and the orbital period $(P_{\rm orb})$, at a rate of $\dot{x}\simeq 2.6\times 10^{-13}\,\text{lt-s}\,\text{s}^{-1}$ and $\dot{P}_{\rm orb}\simeq 4 \times 10^{-10}\,\text{s}\,\text{s}^{-1}$, respectively. \swiftj{} is the only accreting millisecond X-ray pulsar, so far, from which the orbital period derivative has been directly measured from appreciable changes on the observed orbital period.   
Finally, no significant secular deceleration of the spin frequency of the compact object is detected, which allowed us to set a constraint on the magnetic field strength at the polar caps of $B_{PC}<1.3\times 10^{8}~\text{G}$, in line with typical values reported for AMXPs.  
\end{abstract}




\section{Introduction}

More than two decades after the discovery of SAX J1808.4$-$3658 \citep{Wijnands:1998vk}, accreting millisecond X-ray pulsars (hereafter AMXPs) remain key astrophysical objects that give us unique access to investigate extreme physics. AMXPs are rapidly rotating neutron stars (NS) that form tight binary systems with sub-Solar companion stars, characterising them as low-mass X-ray binaries. Their Gyr age combined with the short spin period ($<30$ ms) and low magnetic fields ($<10^9$~G) have been successfully explained by the so-called \textit{recycling scenario} \citep[see e.g.,][]{Alpar:1982wh,Bhattacharya91}. This model describes AMXPs as the final stage of a long-term accretion process in which old slow-spinning radio pulsars are spun-up to millisecond periods by accreting matter and angular momentum transferred via Roche-lobe overflow from a companion star. Coherent X-ray pulsations are expected to arise from the magnetic confinement of the accreting matter that forces the emission of the gravitational binding energy into limited regions of the stellar surface near the magnetic poles, creating bright hot-spots.  
X-ray pulsations are detected during the outburst phases lasting typically between weeks and months \citep[see e.g.,][for reviews]{Di-Salvo:2020va,Patruno:2021vs}.
When the accretion process stops, the fast rotating NS enters quiescence for durations of years to decades. During this period it is expected to shine in the radio as a rotation-powered pulsar. This scenario has been confirmed with the discovery of the AMXP IGR J18245$-$2452 \citep{Papitto:2013uf}, the only source observed to alternate between rotation and accretion powered pulsations. Low luminosity systems such as PSR J1023$+$0038 \citep[see e.g.,][]{Stappers:2014wn,Archibald:2015vw} and XSS J12270$-$4859 \citep[see e.g.,][]{Bassa:2014uu,Papitto:2015wo} showed indirect evidence for such alternating behaviour suggesting similarities with IGR J18245$-$2452. Together they were identified as a sub-class dubbed transitional millisecond pulsars \citep[see][for a review]{Papitto:2020wl}. Indirect evidences for NSs switching between rotation and accretion powered pulsations have been reported also for some AMXPs, on the basis of their quiescent luminosity \citep[see e.g.,][]{Burderi:2003wk,Campana:2004vf,DAvanzo:2009vs}.

Whether or not all the AMXPs behave as rotation-powered pulsars in between outbursts is still difficult to assess based on the observational evidence. However, all the efforts spent so far on searching for radio pulsations from AMXPs in quiescence did not produce encouraging results \citep[see e.g.,][]{Burgay:2003va,Iacolina:2010tn,Patruno:2017ug,Sanna:2018td}. To explain the lack of radio pulsations, \citet{Burderi:2001wp} proposed the so-called \textit{radio-ejection} model, that predicts the activation of the rotation-powered pulsar concurrently with the drop of the mass-transfer rate. Moreover, the model suggests that the radiation pressure generated by the radio pulsar could be large enough to push away from the system the accreting matter transferred via Roche-lobe overflow. If correct, a large fraction of overflowing matter, in the form of strong winds, should engulf the surroundings of the binary system increasing the probability for radio signals generated by the NS to be captured by free-free absorption processes, resulting in a strong reduction or even a total suppression of the original signal. The observation of fast orbital expansion in AMXPs \citep[see e.g.,][]{di-Salvo:2008uu,Burderi:2009td,Sanna:2016ty,Sanna:2017vj} as well as in slowly rotating pulsars or eclipsing LMXBs \citep[see e.g.,][]{Burderi:2010tk,Mazzola:2019wo,Iaria:2018tq} has been interpreted as indirect evidence in favour of this model \citep[however, see e.g.,][for alternative explanations]{Hartman:2008uj,Patruno:2012tw,Patruno:2017ug}. Other indirect evidences corroborating the hypothesis of non-conservative mass transfer in AMXPs have been suggested by \citet{Marino:2019vq} while comparing the observed X-ray luminosities with respect to the values predicted by conservative mass-transfer driven by angular momentum loss via gravitational wave emission and/or magnetic braking. 
The transient behaviour of AMXPs, combined with the typical long-lasting duration of their quiescence phases, represents a crucial limiting factor in the investigation of the long-term evolution of these systems. For only nine AMXPs pulsations have been detected in more than one outburst, however, only for a small subset of them has it been possible to set tight constraints on their orbital evolution.

\swiftj{} was observed for the first time on 2006 June 2 by \swift{}/BAT \citep{Schady2006} triggered by a type-I X-ray burst. A follow-up \swift{}/XRT observation confirmed the nature of the event allowing the identification of the compact object in the binary system and localising the source at a distance $d=6.7\pm1.3$~kpc \citep[however, no sign of photospheric radius expansion was detected for this type-I burst, therefore this value should be considered as an upper limit][]{Wijnands:2009wa,Campana:2009va}. Archival \xmm{} observations allowed the detection of faint X-ray activity from the source at least six years earlier than the detection of the type-I burst. 
\inte{} and \swift{} detected again the source in outburst on 2010 April 10 \citep{Pavan2010,Chenevez2010}. A few days later \rxte{} started a monitoring campaign that revealed the presence of coherent X-ray pulsations at 518~Hz \citep{Altamirano2010} and eclipses at an 8.8~hours periodicity in the X-ray light curve \citep{Markwardt2010}, which makes \swiftj{} the only known eclipsing AMXP. Accurate temporal studies of the X-ray pulsation and eclipses enabled the determination of a precise ephemeris for the NS, setting a tight constraint on the inclination of the binary system and the donor mass in the range 74.4$^\circ$--77.3$^\circ$ and 0.46--0.81~M$_\odot$, respectively, assuming a NS mass in the range 0.8--2.2~M$_\odot$ \citep[][]{Markwardt:2010tl,Altamirano:2011uq}. The mean density value obtained under the assumption that the companion star fills its Roche lobe during the outburst phase suggests a $G3\,V$--$G5\,V$ spectral type if the companion is close to the lower main sequence. No optical counterpart of the system has been detected likely due to the large extinction associated with the measured interstellar hydrogen column density $N_H\sim3 \times 10^{22}\,\text{cm}^{-2}$ \citep{Ferrigno:2011wz}. Searches in the near-infrared (H band) with the VLT/NACO instrument focused on the 1.6 arcsec \swift{} error circle led to the detection of more than 40 possible counterparts \citep{DAvanzo:2011tq}. Based on more accurate localisation (0.6 arcsec 90\% c.l.) obtained by observing the source in quiescence with \chandra{}, further near-infrared observations have been performed with the Gemini-North telescope (k band), limiting the potential counterpart to two candidates \citep{Jonker:2013wp}. 
On 2021 March 1, \inte{}/JEM-X detected X-ray activity, as well as a type-I burst, in the direction of the source, suggesting the onset of a new outburst after eleven years of quiescence \citep{Mereminskiy2021}. Follow-up \nicer{} observations recovered both X-ray pulsations at 518~Hz and eclipses in the light curve, confirming the beginning of a new outburst of \swiftj{} \citep{Bult:2021wk}.

In this work, we carried out the temporal analysis of the latest outburst of the source, combining data collected with \nicer{} and \xmm{}. We then updated the source ephemerides and we discussed the long-term evolution of the orbital parameters in between the two observed outbursts detected so far. Dedicated works on the analysis of the spectral properties of the source (Marino et al. 2022, submitted), the X-ray eclipses (Mancuso et al. 2022, in prep.), and the type-I bursts (Albayati et al. 2022, in prep.) will be following this work.

\section{Observations and data reduction}

\subsection{NICER}

NICER \citep{Gendreau:2012vf} observed the X-ray transient \swiftj{} between 2021 March 1 (MJD 59274.6) and 2021 May 1 (MJD 59335.7) for a total exposure time of almost 160~ks (after applying standard filtering). Here we report on the dataset collected up to 2021 March 13 (MJD 59286.6) in correspondence with the detection of the X-ray pulsations. Due to a visibility gap, the monitoring of the source was interrupted for $\sim2.5$ days in the middle of the outburst. We processed the NICER observations with the \textsc{NICERDAS} pipeline version 7.0 (version V007a) retaining events in the 0.2--12 keV energy range, for which the pointing offset was <54 arcsec, the dark Earth limb angle was $>30^\circ$, the bright Earth limb angle was $>40^\circ$, and the ISS location was outside of the South Atlantic Anomaly (SAA). Moreover, we selected events from 52 out of the 56 aligned pairs of X-ray concentrator optics and silicon drift detectors. Background count rate was extracted for each observation using the \textsc{nibackgen3C50} tool \citep{Remillard:2021tt}. We reported photon arrival times to the Solar System barycentre by using the \textsc{barycorr} tool (DE-405 solar system ephemeris).   


\subsection{XMM-Newton}

XMM-Newton \citep{Jansen2001} triggered a target of opportunity observation of \swiftj{} (Obs. ID. 0872392001) starting from 2021 March 4 at 08:41 UTC and ending on 2021 March 4 at 16:17 UTC, for a total duration of 56.7~ks. 
The instrumental set-up during the observation included the Epic-pn (PN) camera operated in timing mode, the Epic-MOS 1--2 in small window and timing mode, respectively, and the RGS in spectroscopy mode. For the purpose of this work, we focused on the PN observation only. We extracted the PN dataset using the Science Analysis Software (SAS) v.19 with up-to-date calibration files. We retained events in the energy range 0.5–10 keV, by selecting optimally calibrated events with \textsc{PATTERN $\leq$ 4} and \textsc{(FLAG = 0)}. We filtered source and background events from the PN instrument within the RAWX regions [29:47] and [2:8], respectively. We investigated for possible high background flaring activity by constructing a 20s resolution light curve of the source events at energies larger than 10 keV, but we found none.  
The 0.5--10 keV light curve of the source shows an average count-rate of around 40 counts per second. Six type-I X-ray bursts almost evenly spaced with a recurrence time of 2.2 hours, and two full X-ray eclipses are also clearly visible in the light curve. We reported photon arrival times to the Solar System barycentre by using the \textsc{barycen} tool (DE-405 solar system ephemeris)

\subsection{RXTE}

Intending to coherently compare the temporal properties of the two observed outbursts of \swiftj{}, we re-analysed the data collected by the \rxte{}/Proportional Counter Array \citep[PCA; 2--60 keV;][]{Jahoda:2006uf} between 2010 April 14 and 2010 April 20 (Obs. Id. P95085-09). To perform the timing analysis, we used data collected in the event mode characterised by a time resolution of 122 $\mu$s and energy range combined into 64 independent channels. Following \citet{Markwardt:2010tl}, to maximise the signal-to-noise ratio, we retained photons in the energy range 2--30 keV (PCA channels 5--80) collected by all the active PCUs during each of the observations. We reported photon arrival times to the Solar System barycentre by using the \textsc{faxbary} tool (DE-405 solar system ephemeris).

\section{Data analysis and results}
\label{sec:results}

\begin{figure*}
\centering
\includegraphics[width=0.8\textwidth]{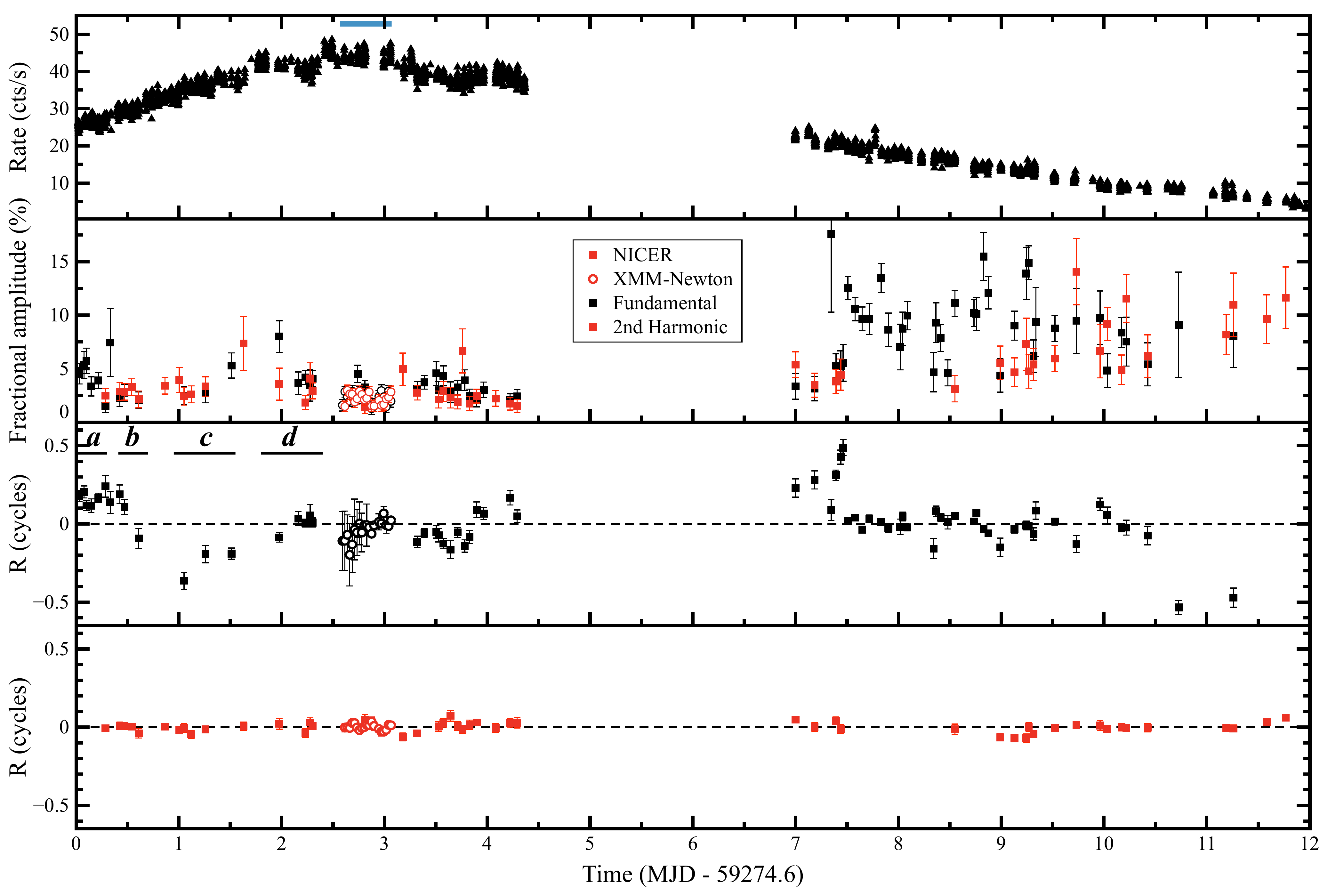}
\caption{\textit{First panel -} \nicer{} 0.5--10 keV light curve of the latest outburst of the accreting millisecond X-ray pulsar \swiftj{} starting from March 1, 2021 (MJD 59274.1). Each points represents the count rate estimated collecting 32s of exposure time. The lack of data in the range 4.5--7 days reflects a visibility gap during the monitoring campaign. The cyan segment identifies the time interval in which the \xmm{} observation has been performed. \textit{Second panel -} Time evolution of the fractional amplitude determined for the fundamental (black points) and the second harmonic (red points) components used to model the source pulse profiles created from the \nicer{} (filled squares) and \xmm{} (empty circles) datasets. \textit{Third panel -} Fundamental pulse phase residuals in units of phase cycles relative to the best-fitting solution. Letters from \textbf{a} to \textbf{d} represent the time intervals selected to investigate the pulse profile shape evolution around a phase jump episode. \textit{Fourth panel -} Second harmonic pulse phase residuals in units of phase cycles relative to the best-fitting solution.}
\label{fig:phase_fit}
\end{figure*}

Both NICER and XMM-Newton observations registered several type-I X-ray bursts and X-ray eclipses during the outburst of the source. To investigate the temporal behaviour of the X-ray pulsations we decided to exclude type-I burst and eclipses from the analysis by ignoring time intervals of $\pm 50$ seconds around the events. The top panel of Figure~\ref{fig:phase_fit} shows the outburst light curve extracted from the \nicer{} dataset. The cyan segment represents the time at which the \xmm{} observation has been performed. Moreover, we retained only events with energies within the range 0.5 to 10 keV. Barycentric correction was performed by using the most accurate coordinates of the source \citep{Jonker:2013wp}.

Starting from the timing solution obtained for the first outburst of the source \citep{Markwardt:2010tl}, we applied barycentric correction to the photon time of arrivals of the \nicer{} and \xmm{} datasets to account for the Doppler effect in the binary system assuming a circular orbit \citep[see e.g.,][for a detailed description of the method]{Burderi:2007tl,Sanna:2016ty}. Following the preliminary detection of the X-ray pulsation during the latest outburst of the source \citep{Bult:2021wk}, we updated the time of passage from the ascending node ($T_{ASC}$) to be able to recover the coherent signal. Before applying phase-coherent timing analysis, we refined the values of the local $T_{ASC}$ and the mean spin frequency. We explored the local $T_{ASC}$ value, by sampling the interval $59274.49427\pm P_{orb}/2$ at 1-second steps. For each $T_{ASC}$ value we applied barycentric correction keeping the other orbital parameters fixed. We then performed epoch-folding searching techniques of the whole dataset using 16 phase bins by exploring the frequency space around the value reported by \citet{Bult:2021wk} with steps of $10^{-8}$~Hz for a total of 10001 steps. Under the assumption that the best orbital solution is represented by the folded pulse profile at the largest $\chi^2$ value in the epoch-folding search with respect to a constant distribution of photons \citep[see
e.g.,][]{Kirsch:2004tg}, we obtained $T_{ASC}=59274.49428$~MJD and $\overline{\nu}=517.92001388$~Hz.
It is worth noting that the estimated local $T_{ASC}$ value significantly differs from the predicted value obtained by propagating the orbital solution reported by \citep{Markwardt:2010tl} assuming a constant orbital period. As will be discussed in detail in the next sections, the latter result suggests the presence of a significant evolution of the orbital period of the system.

To perform coherent timing analysis, we applied the updated local timing solution to 8-phase-bin pulse profiles created by epoch-folding 2000 s-long data segments at the mean spin frequency $\overline{\nu}$. If required, we adjusted the length of the data segments to improve the significance of the pulse profiles. We modelled the pulse profiles by fitting a constant plus a superposition of two harmonically related sinusoidal components. We retained only harmonics for which the ratio between the sinusoidal amplitude and the corresponding $1\sigma$ uncertainty is equal or greater than three. For each significant component, we inferred two quantities: a) the fractional amplitude defined as the ratio between the sinusoidal amplitude and the source photons collected to create the pulse profile (i.e., the total number of photons in the profile minus the corresponding background), b) the fractional part of the phase residual. Higher harmonics were not required to describe the pulse profile on such a relatively short timescale. 

The second panel of Figure~\ref{fig:phase_fit} shows the evolution of the background-corrected fractional amplitude of the fundamental (black) and second harmonic (red) components of the \nicer{} (filled squares) and \xmm{} (empty circles) datasets during the outburst. 

We modelled the fundamental and second harmonic components describing the pulse profiles of the source by applying standard timing techniques \citep[see e.g.,][for more details]{Burderi:2007tl, Sanna:2018wh}. More specifically, we modelled the pulse phase delay temporal evolution as:
\begin{equation}
\label{eq:fit_phase}
\Delta \phi(t)= \Delta \phi_0 - \Delta \nu_0(t-T_0)- \frac{1}{2} \dot{\nu}(t-T_0)^2+R_{orb}(t),
\end{equation}
with $\Delta \phi_0$, $\Delta \nu_0$ and $\dot{\nu}$ representing the pulse phase delay at $T_0$, the correction factor on the frequency used to epoch-fold the data, and the spin frequency derivative referred to the epoch $T_0$. $R_{\rm orb}(t)$ describes the residual orbital modulation caused by the differential corrections between the guessed NS ephemeris and the real one \citep[see e.g.,][]{Deeter:1981te}.

We iteratively repeated this process for each refined ephemeris until no significant improvements were found for any of the model parameters. In Table~\ref{tab:solution} we list the orbital best-fit parameters as well as the spin frequency parameters obtained independently from the analysis of the fundamental and second harmonic components. In the third and fourth panels of Figure~\ref{fig:phase_fit}, we show the phase delay residuals with respect to the best-fitting models for the fundamental and the second harmonic, respectively. 

\begin{table*}

\begin{tabular}{l | c  c  c  c}
\hline
&  2010 & \multicolumn{2}{c}{2021} \\
\hline
Parameters   &    Second Harmonic      & Fundamental & Second Harmonic\\
\hline
\hline
R.A. (J2000) &  \multicolumn{3}{c}{$17^h49^m31^s.73\pm0.04^s\!$}\\
Decl. (J2000) & \multicolumn{3}{c}{$-28^\circ08'05''.08\pm0.6''$} \\
Orbital period $P_{\rm orb}$ (s) & 31740.702(13) & 31740.84(1) & 31740.8417(27)\\
Projected semi-major axis $x$ (lt-s) & 1.89948(2) & 1.89956(3) &1.899568(11) \\
Ascending node passage $T_{{\rm ASC}}$ (MJD) & 55300.6522518(17) & 59274.494176(5) & 59274.4941787(14)\\
Eccentricity ($e$) & $4.3(1.3)\times 10^{-5}$ & $3.7(3.3)\times 10^{-5}$ &$4.1(1.1)\times 10^{-5}$\\
$\chi^2$/d.o.f. & 165.0/42 & 1001.6/84 & 97.8/60\\
\hline
\hline
Spin frequency $\nu_0$ (Hz) & 517.92001375(23)$^*$ &517.92001572(25)$^*$& 517.92001385(16)$^*$\\
Spin frequency 1st derivative $\dot{\nu}_0$ (Hz/s) & 6.5(2.5)$\times 10^{-12}$$^*$ &-4.0(5)$\times 10^{-12}$$^*$&$-0.6(1.1)\times 10^{-13}$$^*$\\
\hline
\end{tabular}
\caption{Timing solutions obtained from the analysis of the \rxte{} observations of the 2010 outburst and the \nicer{} and \xmm{} observations collected during the whole 2021 outburst of \swiftj{}. The orbital solutions are  referred to the epochs T$_0$=59274.6~MJD and T$_0$=55300.0~MJD, respectively. Errors are at 1$\sigma$ confidence level. The reported X-ray position of the source has been determined by observing the source in quiescence with \chandra{} \citep{Jonker:2013wp}.$^*$Uncertainties have been calculated including the contribution of the positional uncertainties.}
\label{tab:solution}
\end{table*}

To take into account large values of the reduced $\chi^2$ ($\tilde{\chi}^2>1$), we rescaled the uncertainties on the parameters reported in Table~\ref{tab:solution} by a factor $\sqrt{\tilde{\chi}^2}$ \citep[see e.g.,][]{Finger:1999vb}.

We investigated the contribution of the positional uncertainties on the spin frequency and its derivative by considering the residuals induced by the Earth's motion assuming small variations of the source position $\delta_{\lambda}$ and $\delta_{\gamma}$ (ecliptic coordinates) expressed by the relation:
\begin{equation}
R_{pos}(t) = - \nu_0 y [\sin(M_0+\epsilon)\cos \gamma \delta\lambda -  \cos(M_0+\epsilon)\sin \gamma \delta\gamma],
\label{eq:pos}
\end{equation}
where $y=r_E/c$ represents the Earth semi-major axis in light-seconds, $M_0=2 \pi (T_0-T_{v})/P_{\oplus}-\lambda$, $T_{v}$ and $P_{\oplus}$ are the vernal point and the Earth orbital period, respectively, and $\epsilon=2\pi(t-T_0)/P_{\oplus}$ \citep[see, e.g.,][]{Lyne90}. Given the short duration of the outburst with respect to Earth's orbital period, we can solve Eq.~\ref{eq:pos} by expanding it in a series of $\epsilon\ll1$ \cite[see e.g.,][and references therein]{Burderi:2007tl}. This allows estimating upper limits on the spin frequency correction and the spin derivative as $\sigma_{\nu_{\rm pos}}\leq \nu_0y\sigma_{v}(1+\sin^2\gamma)^{1/2}2\pi/P_{\oplus}$ and $\sigma_{\dot{\nu}_{\rm pos}}\leq \nu_0y\sigma_{v}(1+\sin^2\gamma)^{1/2}(2\pi/P_{\oplus})^2$, respectively, where $\sigma_{v}$ represents the positional error circle. Using the source position reported in Table~\ref{tab:solution} \citep{Jonker:2013wp}, we estimated $\sigma_{\nu_{\rm pos}} \leq 1.4\times 10^{-7}$~Hz and $\sigma_{\dot{\nu}_{pos}} \leq 3\times 10^{-14}~\text{Hz}~\text{s}^{-1}$, respectively. We added in quadrature these systematic uncertainties to the statistical errors of $\nu_0$ and $\dot{\nu}$ estimated from the timing analysis.

To consistently investigate possible variations on the temporal properties of the NS obtained from the analysis of the two observed outbursts of \swiftj{}, we re-analyse the \rxte{} observations performed during its 2010 outburst. As mentioned above, a more accurate set of coordinates has been obtained during the quiescence phase that followed the first outburst of the source \citep{Jonker:2013wp} as well as the first accurate orbital ephemerides estimation \citep{Markwardt:2010tl}. Although we do not expect significant discrepancies on the inferred orbital parameters, variations on the spin frequency and its derivative are predicted due to the residual effect from the barycentric correction discussed earlier. Following the procedure discussed for the analysis of the \nicer{} and \xmm{} datasets, we estimated the pulse phase delays from 8-phase-bin pulse profiles obtained by epoch-folding 500 s-long data segments using the timing solution reported by \citet{Markwardt:2010tl} as a starting point. We then fitted separately the fundamental and second harmonic phase delays with the model described in Eq.~\ref{eq:fit_phase}. In Table~\ref{tab:solution} we report the orbital best-fit parameters, as well as the spin frequency parameters obtained from the analysis of the second harmonic since the presence of a large phase jump ($\sim0.4$ cycles) in the fundamental component significantly reduced the accuracy of the associated timing solution \citep[a similar conclusion has been reached by][]{Markwardt:2010tl,Altamirano:2011uq}. As expected, the orbital solution obtained from our analysis is consistent within errors with that reported by \citet{Markwardt:2010tl}, however, we note discrepancies in the accuracy achieved in the two analyses, which are likely associated with the different methods applied.

Finally, by combining orbital ephemerides reported in Table~\ref{tab:solution} for the two outbursts, we investigated the long-term orbital evolution by studying independently the variation of the orbital period and the delay accumulated by $T_{\rm ASC}$ as a function of the orbital cycles elapsed since its discovery. 

The variation of the orbital period between the two outbursts of \swiftj{} is $\Delta P_{\rm orb}=0.140(13)$~s. Interestingly, and not commonly reported for other AMXPs, the secular orbital evolution for this source can be directly estimated by comparing the orbital period values thanks to the relatively large variation and the small uncertainties associated with the single measurements (something similar has been observed for the AMXP IGR J17511-3057; Riggio et al. 2022, in prep.). Taking into account that the number of orbital cycles elapsed in between the two outbursts is $N=10817$, we can estimate an orbital period derivative $\dot{P}_{\rm orb}=4.1(3)\times 10^{-10}\,\text{s}\,\text{s}^{-1}$ compatible with a fast expansion of the binary system.

The presence of a significant first derivative of the orbital period of the system should affect the evolution of other orbital parameters such as the time of passage of the ascending node. In fact, by extending the analogy of coherent timing analysis to the orbital parameters, we can approximate $T_{\rm ASC}$ with the expression:
\begin{equation}
\label{eq:evo}
T_{\rm ASC}=T_{0,{\rm ASC}}+N P_{0,{\rm orb}}+\frac{1}{2}N^2P_{0,{\rm orb}}\dot{P}_{\rm orb},
\end{equation}
where $T_{0,{\rm ASC}}$, $P_{0,{\rm orb}}$ and $\dot{P}_{\rm orb}$ represent the time of passage of	the ascending node, the orbital period and its derivative with respect to a specific reference epoch, respectively. 
To verify the compatibility of the $T_{\rm ASC}$ values measured in the two outbursts with that inferred from Eq.~\ref{eq:evo}, we take the 2010 orbital solution as a reference and we estimated the $T_{\rm ASC}$ value for the latest outburst assuming $\dot{P}_{\rm orb}=4.1(3)\times 10^{-10}~\text{s}~\text{s}^{-1}$. To be able to use Eq.~\ref{eq:evo}, we quantify the number of orbital cycles elapsed in between the two outbursts as N=INT[($T_{2021,{\rm ASC}}-T_{2010,{\rm ASC}}$)/$P_{2010,{\rm orb}}$], where INT represents the closest integer. Combining all, we estimate a guess for the time of passage from the ascending node of $T_{2021,{\rm ASC}}=59274.4950(14)$~MJD, which is compatible within uncertainties with the measured value reported in Table.\ref{tab:solution}.

\section{Discussion}

We have reported an updated timing solution for the peculiar accreting millisecond X-ray pulsar \swiftj{} obtained by performing phase-coherent timing analysis of the X-ray pulsations detected by \nicer{} and \xmm{} during its latest outburst. Interestingly, the new set of orbital parameters shows significant signs of temporal evolution with respect to the timing solution obtained from the first outburst \citep[see also][]{Markwardt:2010tl,Altamirano:2011uq,Ferrigno:2011wz}, finding for the first time significant variations on both the orbital period ($\Delta P_{\rm orb}$), and the NS projected semi-major axis ($\Delta x$). Moreover, at odds with other AMXPs, we find marginally significant evidence for non-zero eccentricity. In the following, we will discuss each of these parameters in detail, however, in broad terms we emphasise that the large orbit as well as the high inclination of \swiftj{} with respect the the AMXPs, might be the reason why we reached higher sensitivity on its orbital parameters.  


\subsection{Phase and pulse profile evolution}
\label{sec:phase}
The first thing that can be noticed by visually inspecting the evolution of the pulse profile phases reported in the third panel of Figure~\ref{fig:phase_fit} is the clear presence of phase jumps in the fundamental harmonic of the signal. This phenomenon is evident around days 1, 7, and 11 with respect to the beginning of the monitoring campaign. These jumps are nearly half pulse phase cycle on time-scale of a few days, implying unrealistic local spin frequency derivatives of the order of $\dot{\nu} \approx 10^{-10}$ Hz s$^{-1}$ or huge longitudinal movements of the hot spot of the order of the NS radius, as we will discuss later. Interestingly, no evidence of this phenomenon is shown by the temporal evolution of the second harmonic (Figure~\ref{fig:phase_fit}, fourth panel). Similar behaviour is observed also during the 2010 outburst, where the fundamental phase shows a drop of $\sim0.4$ \citep[see e.g.,][]{Altamirano:2011uq}. Phase shifts of the fundamental component up to 0.3~cycles have been already reported in three more AMXPs, i.e., SAX J1808.4$-$3658 \citep{Burderi:2006va}, XTE J1814$-$338 \citep{Papitto:2007wp}, and XTE J1807$-$294 \citep{Riggio:2008wz,Patruno:2010wi}. Moreover, by comparing the third and the first panels of Figure~\ref{fig:phase_fit}, no clear correlation between phase jumps and X-ray count rate (assumed as a good proxy of the source flux) seems to be present unlike some of the systems previously mentioned \citep[see e.g.,][]{Papitto:2007wp,Patruno:2010wi}. It is worth noting that sources showing phase jumps present pulse profiles with the richest harmonic content, while on the other hand, standard AMXPs show profiles well described by a sinusoidal function. From the evolution of the fractional amplitude of the two components used to describe the pulse profile (Figure~\ref{fig:phase_fit}, second panel), we can notice that both the fundamental and the second harmonic vary significantly during the outburst. More specifically, during the first four days of the outburst, the fractional amplitude of the fundamental is approximately constant around the value 3.2\%, while after the observational gap the mean value increases up to 8.4\% and a large scatter around this value is visible. Quite similarly, the fractional amplitude of the second harmonic is almost constant around the value of 2.5\% for the first part of the outburst, reaching up to 6.8\% in the second part with a significant increasing trend. Finally, it is noteworthy that jumps in the phase of the fundamental component seem to be happening when the second harmonic fractional amplitude is larger than (or at least comparable to) the fundamental.

\begin{figure}
\centering
\includegraphics[width=0.5\textwidth]{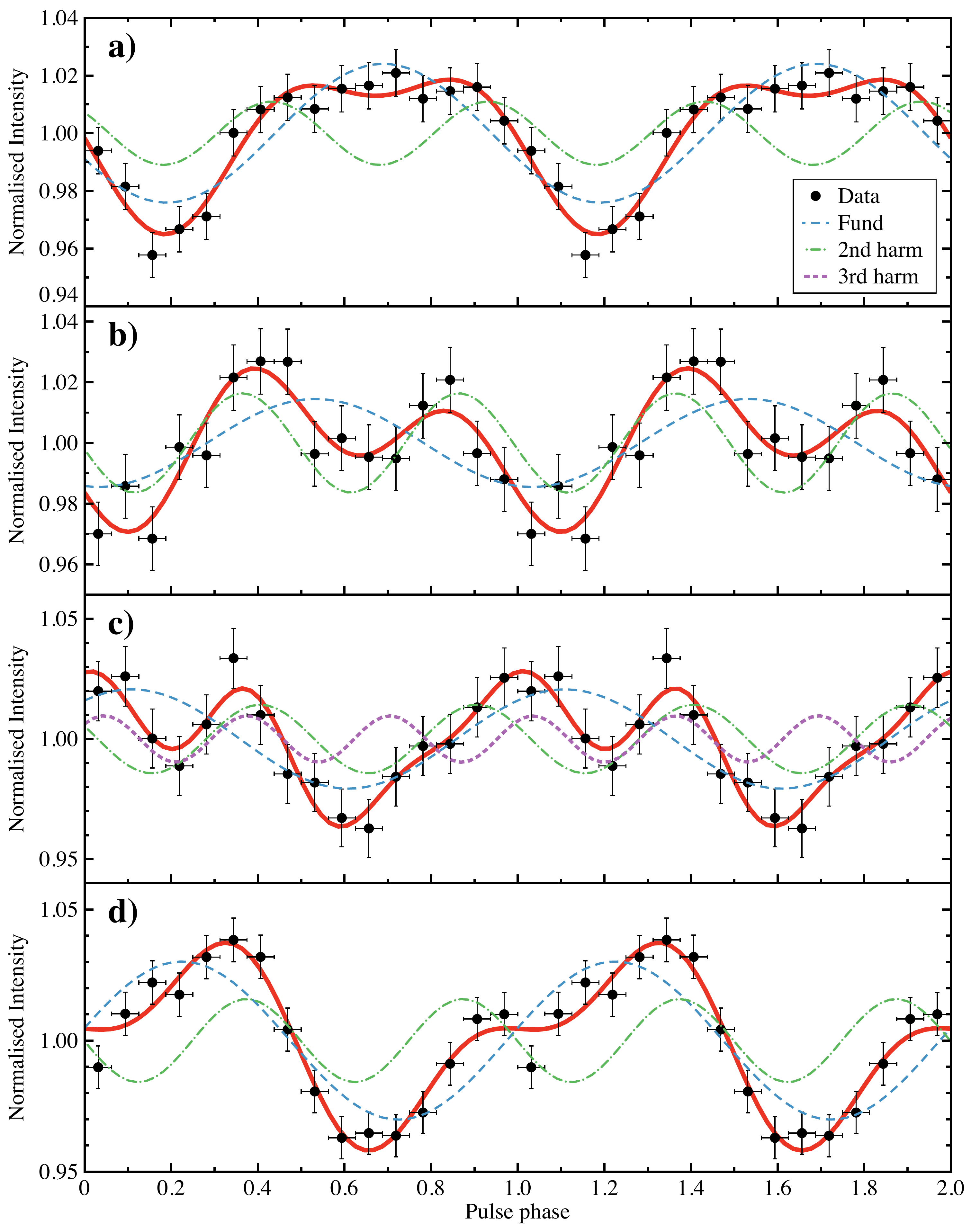}
\caption{Evolution of the X-ray pulse profile during the first $\sim 2.5$~days of the \nicer{} monitoring campaign of \swiftj{}. The four panels show the pulse profiles (black points) obtained by collecting events into four time intervals (highlighted with black lines and labelled from \textbf{a} to \textbf{d} in Figure~\ref{fig:phase_fit}) around the phase jump of the fundamental component. The best-fitting models (red lines) are the superposition of two (first and fourth panels) and three (second and third panels) harmonically related sinusoidal functions. For clarity, two cycles of the pulse profile are shown.}
\label{fig:profile_evo}
\end{figure} 

To further investigate this phenomenon, we isolated the first phase jump as a representative example, and we explored the pulse profile evolution in the restricted time interval including $\sim2.5$~days since the beginning of the \nicer{} campaign. More specifically, we divided the interval into four segments that we identify with letters from a) to d), as reported in the third panel of Figure~\ref{fig:phase_fit}. From top to bottom of Figure~\ref{fig:profile_evo}, we show the \nicer{} pulse profiles obtained by correcting the photons for the updated timing solution reported in Table~\ref{tab:solution}. Significant changes in the pulse shape are observed in correspondence with the time evolution of the drift of the fundamental pulse phase. Two interesting aspects can be noticed, firstly at the onset of the phase jump (Figure~\ref{fig:profile_evo}, panel b) the second harmonic overcomes the fundamental becoming the dominant component; secondly, the pulse profile corresponding to the largest phase jump amplitude increases its harmonic contents with the appearance of a relatively intense third harmonic component (Figure~\ref{fig:profile_evo}, panel c). Similar behaviour is observed from the analysis of the 2010 outburst of the source, where the pulse profiles near the unique phase jump detected significantly require additional third and fourth harmonics to be modelled \citep{Altamirano:2011uq}. Interestingly, as the phase recovers part of its shift, the profile loses the higher harmonic content and the fundamental component is again predominant (Figure~\ref{fig:profile_evo}, panel d) as in the pre-jump interval (Figure~\ref{fig:profile_evo}, panel a), although with significantly different pulse form due to the variation of the relative phase shift of the two components.

The origin of phase fluctuations, as well as phase jumps in these systems, are still an open question, however, several possible mechanisms have been proposed over the years. Given the direct correlation between the pulse phase and the NS rotation period, it is easy to associate phase variations with mere changes in the NS spin rate. However, AMXPs showed jumps of the order of several tenths of the pulse phase on timescales of days, implying large spin derivatives difficult to explain with standard known mechanisms. Moreover, since jumps have been observed in both directions, we need to invoke very powerful spin-up and spin-down alternatives to standard accretion torques \citep[see e.g.,][]{Lamb:1978uf,Ghosh:1979ws}, magnetic torques \citep[see e.g.,][]{Wang95,Rappaport:2004tr,Kluzniak07}, or spin-down torques due to emission of gravitational radiation \citep[see e.g.,][]{Bildsten:1998wb}.
Another aspect that could be taken into account is the interplay between the magnetic field and accreting matter on the surface of the NS that generates the hot spot responsible for the observed X-ray modulation. Slow movements of the location of either the magnetic poles or the matter deposition point near the magnetic poles would cause pulse phase changes. In the former case, processes such as localised magnetic burial (or screening) and amplification should be invoked on timescales as short as days. However, current understanding indicates that a significant reduction of the NS magnetic dipole moment can be achieved by accreting at least $10^{-5}$~M$_\odot$ \citep{Payne:2004wj}, which makes this process inefficient on short timescales taking into account that AMXPs typically accrete at most a few percent of the Eddington limit \citep[see e.g.,][]{Di-Salvo:2020va,Patruno:2021vs}. In the latter case, simulations of funnelled accretion in which hot-spots are close to the spin axis (scenario expected for small misalignment angles between the NS magnetic poles and the spin axis) and move in the azimuthal direction, clearly suggest that small displacements can produce phase shifts up to several tenths of a cycle \citep[see e.g.,][and references therein]{Lamb:2009wf}.     
\citet{Poutanen:2009wb} suggested that timing noise observed in AMXPs derives from the evolution of the pulse form during the outburst. By modelling the pulse profile accounting for the visibility of the two antipodal hot-spot as a function of the geometrical size of the accretion disk, these authors argue that sudden phase shifts can be explained with an increase of visibility of the antipodal spot caused by the receding accretion disk \citep[see e.g., the 2002 outburst of SAX J1808.4$-$3658;][]{Poutanen:2009wb}. The same phenomenon has been interpreted as a purely geometrical effect based on the simple principle that the phase of a signal may change if the relative strength of the two polar caps varies. According to this scenario, the observed erratic pulse phase shifts do not reflect a real movement of the X-ray emitting region, but simply a different redistribution of the accretion flow between the two hot-spots \citep[][Riggio et al. 2022, in prep.]{Riggio:2011ua}. Moreover, this model predicts that, when present, a third harmonic should follow the same erratic pulse phase shifts of the fundamental one. Suitable observations of the source with large statistics on short timescales are likely needed to test this hypothesis.

Complexity in the magnetic field structure could as well play a crucial role in causing the phase wandering observed in AMXPs. The canonical sinusoidal pulse profiles detected in these systems are consistent with the standard picture of a dipolar magnetic field slightly tilted with respect to the spin axis, however, similar pulse shapes can arise from objects with more complex fields \citep[see e.g.,][]{Long:2008vb}. Interestingly, in the case of accreting NS characterised by multipolar fields, phase shifts, as well as pulse profile changes, are expected from the interaction between the magnetic field and the accretion disk at different accretion rates. In fact, for increasing accretion rates, the disk moves inwards interacting firstly with the dipolar component (dominant at large distances) and then with higher-order multipoles of the field when it gets closer to the NS surface. A variation in both the pulse form and the phase is expected due to the different magnetic axes associated with the multipoles, which will cause the variation of the hot-spot region at different accretion rates. It is noteworthy that phase jumps due to this process should only be temporary, i.e., the phase slowly comes back to the pre-jump value as the accretion rate returns to its previous value \citep{Long2012}.

\subsection{Eccentricity}
\label{sec:ecc}
Results from the coherent timing analysis performed on the two outbursts of \swiftj{} (see Table~\ref{tab:solution}) suggest a marginally significant detection (slightly above 3~$\sigma$) of the binary eccentricity. The two independent measurements are consistent within errors and in line with results reported by \citet{Markwardt:2010tl} for the 2010 outburst of the source. It is noteworthy that \swiftj{} is currently the only AMXP for which we can constrain this parameter, while for the others only upper limits have been reported.  
To investigate the robustness of the result with respect to bias due to data sampling effects, we performed simulations by applying Jackknife resampling methods \citep[see e.g.,][]{Shao1995}. As a first attempt, we generated 1000 samples starting from the original dataset of \nicer{}/\xmm{} pulse phase delays from which we excluded each time 20\% of the points extracted randomly with respect to the original sample. We then performed coherent timing analysis on each of the simulated datasets and we compared the obtained orbital parameters. The same analysis has been replicated by excluding 30\% and 50\% of the original points. Interestingly, we find that the mean value of the eccentricity inferred from the fit of the simulated samples is always consistent with the estimate obtained from the original dataset, strengthening the reliability of the detection. As predicted, the width of the distribution of the orbital parameters increases as the number of points in the pulse phase datasets decreases.

Even though millisecond pulsars in binary systems, especially those observed in the Galactic plane, are usually associated with nearly circular orbits, a significant fraction of these systems are observed in orbits with values of eccentricity ranging between $\sim 10^{-2}$ to $\sim 10^{-6}$ \citep[roughly 60\% of the binary millisecond pulsars reported in the \textit{ATNF} catalogue meet this condition;][]{Manchester:2005tg}. Current upper limits reported for AMXPs are in line with this distribution. In accordance with the recycling scenario \citep[see e.g.,][]{Bhattacharya91}, millisecond pulsars belong to old binary systems, which makes the previous result difficult to reconcile with the tidal circularisation process that on relatively short timescales is expected to damp orbital eccentricities as high as that acquired by a binary system surviving a supernova explosion. Several mechanisms that can induce eccentricity to circularised orbits have been proposed to try explaining observations, e.g., close encounters between binary systems and near-by passing stars (however, unless the binary system is part of a globular cluster, the probability to have a star passing by close enough to excite its eccentricity values is very low); fast mass-loss episodes at the end of the red giant's evolutionary phase; non-axisymmetric configuration of accretion disks around accreting NS could induce torques able to increase the orbital eccentricity (it should be noted, however, that given the relatively low mass content of typical accretion disks in LMXBs, this effect is expected to be irrelevant) and hierarchical multiple systems. Another very promising process that has been proposed by \citet{Phinney:1992ty}, explains and predicts the low values of eccentricity in these systems as the result of the fluctuation-dissipation theorem applied to the convective eddies in the outer layers of the donor during the last stages of its red giant phase, right before being replaced by a white-dwarf core as a companion of the NS. Interestingly, this prescription seems to match (at least of the order of magnitude) the eccentricities of pulsars bound with low-mass white dwarf companions. However, it is noteworthy that donor stars in AMXPs are likely less massive objects, therefore, it remains to be verify the efficiency of such a mechanism in these systems.

\subsection{Secular evolution properties}

\subsubsection{Orbital period evolution} 
\label{sec:orb}

\swiftj{} has been detected in outburst only twice during the last eleven years. We measured the orbital period to high accuracy in each outburst, finding a significant orbital period variation over these eleven years. More specifically, in between the two outbursts, the orbital period varied by the quantity $\Delta P_{\rm orb}=0.140(13)\,\text{s}$, which corresponds to an orbital period derivative $\dot{P}_{\rm orb}=4.1(3)\times 10^{-10}\,\text{s}\,\text{s}^{-1}$ with respect to the reference epoch $T_0 = 55300.6$~MJD. It is noteworthy that this is the first direct measurement of the orbital period derivative for an AMXP, indeed previously reported estimates of this parameter are based on the temporal drift of the time of passage of the ascending node for systems observed in outburst at least three times.  

Interestingly, as reported in Sec.~\ref{sec:results} we can cross-check the orbital period expansion estimate by verifying its compatibility with the temporal evolution of the time of passage of the ascending node. Since the two methods are independent, the consistency of the two results validates the reliability of the indirect method.  

Currently, orbital period derivatives have been estimated for seven AMXPs out of the twenty-one known including the source subject of this work. Similarly to SAX J1808.4$-$3658 \citep[$\dot{P}_{\rm orb}=1.7(5)\,\times 10^{-12}\text{s}\,\text{s}^{-1}$, see e.g.,][]{di-Salvo:2008uu, Patruno:2017ah,Sanna:2017vj,Bult:2020tu}, \swiftj{} shows a significant long-term fast orbital expansion of the system (although almost two orders of magnitude faster). Fast orbital period expansion ($\dot{P}_{\rm orb}=8.4(20)\,\times 10^{-12}\text{s}\,\text{s}^{-1}$) was observed also for IGR J17062$-$6143 \citep{Bult:2021vs}. For the other systems, uncertainties are such that it is not yet possible to determine whether the orbits are secularly expanding or shrinking. Upper limits on the orbital period derivatives from SAX J1748.9$-$2021 ($\dot{P}_{\rm orb}$ in the range $(-0.7\,\mbox{--}\,8.4)\times 10^{-11}\text{s}\,\text{s}^{-1}$; Sanna et al. 2022, in prep.), SWIFT J1756.9$-$2508 \citep[$\dot{P}_{\rm orb}$ ranging between $(-4.1\,\mbox{--}\,7.1)\times 10^{-12}\text{s}\,\text{s}^{-1}$, see e.g.,][]{Bult:2018ve,Sanna:2018aa}, IGR J17379$-$3747 \citep[$\dot{P}_{\rm orb}$ between $(-9.4\,\mbox{--}\,4.4)\times 10^{-12}\text{s}\,\text{s}^{-1}$;][]{Sanna:2018tx} and XTE J1751$-$305 ($\dot{P}_{\rm orb}$ between $(-2.7\,\mbox{--}\,0.7)\times 10^{-11}\text{s}\,\text{s}^{-1}$; Riggio et al. 2022, in prep.) are still compatible with the fast expansion reported for SAX J1808.4$-$3658, but not with \swiftj{}. On the other hand, a more stringent constraint suggests a much slower evolution for IGR J00291$+$5934 \citep[$\dot{P}_{\rm orb}$ between $(-6.6\,\mbox{--}\,6.5)\times 10^{-13}\text{s}\,\text{s}^{-1}$, see e.g.,][]{Patruno:2017vp, Sanna:2017tx}.

What causes the fast expansion of the orbital period detected in these AMXPs is still unknown, however, over the years several mechanisms have been investigated, although not yet conclusively \citep[see e.g.,][]{di-Salvo:2008uu, Hartman:2008uj,Burderi:2009td, Patruno:2012tw, Sanna:2017vj, Sanna:2018tx}. Nonetheless, a quick estimate of the mass-loss rate from the secondary as a function of the observed orbital period derivative clearly excludes that a standard conservative mass transfer could power such a fast expansion observed for \swiftj{} \citep[similar conclusions have been drawn for SAX J1808.4$-$3658, see e.g.,][]{Sanna:2017vj}. Starting from Kepler's third law, we set the mass transfer condition via Roche-lobe overflow as $\dot{R}_{L2}/R_{L2}=\dot{R}_2/R_{2}$, where the $R_{L2}$ and $R_{2}$ represent the Roche-lobe and secondary radii, respectively. We then express the average long-term secondary mass-loss rate as: 
\begin{eqnarray}
\label{eq:m2_pdot}
-\dot{M}_{2}=\frac{1.94}{(3n-1)}\,m_{2}\Big(\frac{\dot{P}_{{\rm orb},-10}}{P_{{\rm orb},9h}}\Big)\times 10^{-7} \text{M}_{\odot}\, \text{yr}^{-1},
\end{eqnarray}
where -$\dot{M}_{2}$ indicates that the donor star looses matter, $n$ is the stellar index associated with the mass-radius relation of the secondary R$_2\propto$ M$_2^n$, $m_{2}$ is the mass of the companion star in solar mass units, $\dot{P}_{{\rm orb},-10}$ represents the orbital-period derivative in units of $10^{-10}\,\text{s}\,\text{s}^{-1}$, and $P_{{\rm orb},9h}$ is the orbital period in units of 9 hours \citep[see e.g.,][for more details on the derivation of the expression]{Burderi:2010tk}. Eq.~\ref{eq:m2_pdot} shows that mass transfer from the secondary causes an orbital expansion only for secondary stellar indexes $n<1/3$. According to \citet{Markwardt:2010tl}, the donor star has a mass ranging between 0.46~M$_\odot$ and 0.81~M$_\odot$ for a NS in the range 0.8--2.2 M$_\odot$, which likely represents an object laying in the low-mass part of the main sequence branch. These objects are characterised by $n \sim 0.7$ \citep[see e.g.,][]{Chabrier:2009vh}, which implies a decrease of their radius as a response of mass-loss on thermal timescales. Adopting this prescription into Eq.~\ref{eq:m2_pdot}, we reach the conclusion that secondary mass transfer causes the shrinking of the orbit, contrary to what we observe for \swiftj{}. However, inverted mass-radius relations are predicted for main-sequence stars with masses $<1.1$~M$_{\odot}$ undergoing intense mass transfer on timescales shorter than the thermal $\tau_k$ (also known as adiabatic evolution). This is guaranteed by the fact that Sun-like main sequence stars have a surface convective zone sufficiently deep to trigger dynamical instabilities \citep[see e.g.,][]{Ge:2015ug}. As the main-sequence mass decreases, the surface convection zone increases and the adiabatic stellar index approaches the value $n=-1/3$ that characterises fully convective stars \citep[well described with a polytrope index 3/2, see e.g.,][]{Rappaport:1982vc}. 
We can estimate the required secondary mass-transfer rate $\dot{M}_2$ by imposing the mass-loss timescale $\dot{M}_2/|M_2|$ shorter than the thermal timescale $\tau_k=GM_2^2/R_2L_2$, where $M_2$, $R_2$ and $L_2$ represent the mass, radius, and luminosity of the donor star. Assuming that the donor fills its Roche-lobe, we can approximate $R_2\simeq R_{L2}$, where the radius of the Roche-lobe equivalent sphere can be expressed as $R_{L2}\simeq m_2^{1/2} P_{9h}^{2/3}$ R$_\odot$ \citep[obtained by combining Paczy\'nski's Roche-lobe analytic approximation with Kepler's third law;][]{Paczynski71}. The donor luminosity can then be inferred by assuming the main sequence mass-luminosity relation $L_2/L_{\odot}=(M_2/M_{\odot})^{4.5}$ valid for masses in the range 0.5--2 M$_\odot$ \citep{Chabrier:2009vh}. For a donor star in the range 0.46--0.81 M$_\odot$, we find that adiabatic evolution (that allows us to adopt $n\sim-1/3$) requires $\dot{M}_2>(0.2\mbox{--}1.3)\times 10^{-8}~\text{M}_{\odot}~\text{yr}^{-1}$.

Under the hypothesis of adiabatic mass-loss from the donor star ($n=-1.3$), Eq.~\ref{eq:m2_pdot} allows us to infer a mass-transfer rate in the range $\dot{M}_{2}\simeq(-2.1 ~\mbox{--}~ -3.3)\times 10^{-7}~\text{M}_{\odot} \text{yr}^{-1}$. Since the conservative scenario implies that no mass is transferred from the donor during the quiescence phase, Eq.~\ref{eq:m2_pdot} represents a lower limit on the mass transfer rate during the outburst phases. The secular mass transfer inferred from the observed fast orbital expansion of the system would therefore imply a super-Eddington (considering $\dot{M}_{Edd}\sim2.1\times 10^{-8} \text{M}_{\odot}~\text{yr}^{-1}$ for a 1.4~$\text{M}_{\odot}$ NS with radius $\text{R}_{NS}=10$~km) mass transfer rate not compatible with the flux measurements obtained from the observed outbursts of the source. Therefore, this discrepancy strongly suggests that the observed orbital period derivative cannot be explained with a conservative mass-transfer scenario. The same conclusion is confirmed independently from the comparison of the observed and the predicted luminosity values obtained by applying conservative mass-transfer driven by angular momentum loss via gravitational wave emission and/or magnetic braking interactions (Marino et al. 2022, submitted).

\paragraph{Highly non-conservative mass-transfer scenario}
\label{sec:ncmt}
An alternative scenario that can be explored requires the possibility that matter from the companion star is ejected from the binary system, carrying away its angular momentum \citep[see e.g.,][for more details]{di-Salvo:2008uu,Burderi:2009td}. We can parametrize the rate of mass accreting onto the NS, $\dot{M}_1$, as a function of the donor mass-transfer rate $\dot{M}_2$ with the expression $\dot{M}_1=-\beta \dot{M}_2$. As a first hypothesis, we can assume that matter is continuously transferred from the companion star independently from the X-ray activity around the primary object (i.e., both during the outburst and the quiescent phases). We can then express the parameter $\beta$ as the ratio between the observed mass accretion rate derived from the bolometric luminosity extrapolated from the X-ray activity during the outburst phase and the donor mass-transfer rate from Eq.~\ref{eq:m2_pdot}. Considering the peak bolometric luminosity reported during the latest outburst of \swiftj{} (Marino et al. 2022, submitted) $L\sim7.6\times 10^{36}\,\text{erg}\,\text{s}^{-1}$ assuming a 6.7~kpc distance \citep{Wijnands:2009wa}, we can infer a lower limit on the mass-accretion rate on the NS of $\dot{M}_1\simeq6.5\times 10^{-10}{M}_{\odot} \text{yr}^{-1}$. Substituting these values into the expression of $\beta$, we estimate $0.2\%<\beta<0.3\%$. Given the high inclination angle of the source, the bolometric luminosity and the corresponding $beta$ parameter should be considered as lower limits. Moreover, it is noteworthy that such a small value for the $\beta$ parameter implies that almost all matter transferred from the companion star is ejected from the binary system leaving no direct traces of it, except for the fast orbital expansion of the binary system that should represent the only indirect proof of the process. We stress that at this stage, the proposed scenario implies that mass transfer is always active, even in the quiescence phase. We will try to address this point after exploring a suitable configuration of the system  for which the long-term orbital evolution is compatible with a highly non-conservative mass transfer scenario. 
We start by investigating the instantaneous total angular momentum of the binary system and its possible sources of variation in time. Under the hypothesis of a circular orbit, we can express it as:
\begin{equation}
\label{eq:binarymoment}
J_{\rm orb}=\sqrt{\frac{Ga}{(M_1+M_2)}}M_1M_2,
\end{equation}
where $a$ represents the orbital separation, $M_1$ and $M_2$ represent the primary and secondary mass, respectively. By differentiating Eq.~\ref{eq:binarymoment}, it is clear how the evolution of the orbital separation relates to the total rate of change of the binary orbital momentum. Following \citet{di-Salvo:2008uu}, we can describe the evolution of the orbital period as follows:
\begin{equation}
\label{eq:evomom}
\frac{\dot{P}_{\rm orb}}{P_{\rm orb}}=\frac{1}{3}\left[\left( \frac{\dot{J}}{J_{\rm orb}} \right)_{\rm GR}+\left( \frac{\dot{J}}{J_{\rm orb}} \right)_{\rm MB}\right]-\frac{1}{3}\left[\frac{\dot{M}_2}{M_2} g(\alpha,\beta,q)\right],
\end{equation}
where the first term of the right-hand side represents the loss of angular momentum due to emission of gravitational radiation (GR) and magnetic braking (MB) effects, while the second term encapsulates the momentum losses related to the matter ejected by the system through the function $g(\alpha,\beta,q) = 1 - \beta (m_2/m_1) - (1 -(m_2/m_1))(\alpha + q/3)/(1 + q)$, where $m_1$ represents the primary mass in solar mass units. The expression takes into account the mass ratio of the system ($q=m_2/m_1$), the fraction of the donor mass accreting onto the NS ($\beta$) and the specific angular momentum carried away by the ejected matter $\alpha = l_{ej}P_{\rm orb} (m_1 + m_2)^2/(2\pi a^2 m_1^2)$ (expressed with respect to the secondary star).  

\begin{figure}
\centering
\includegraphics[width=0.5\textwidth]{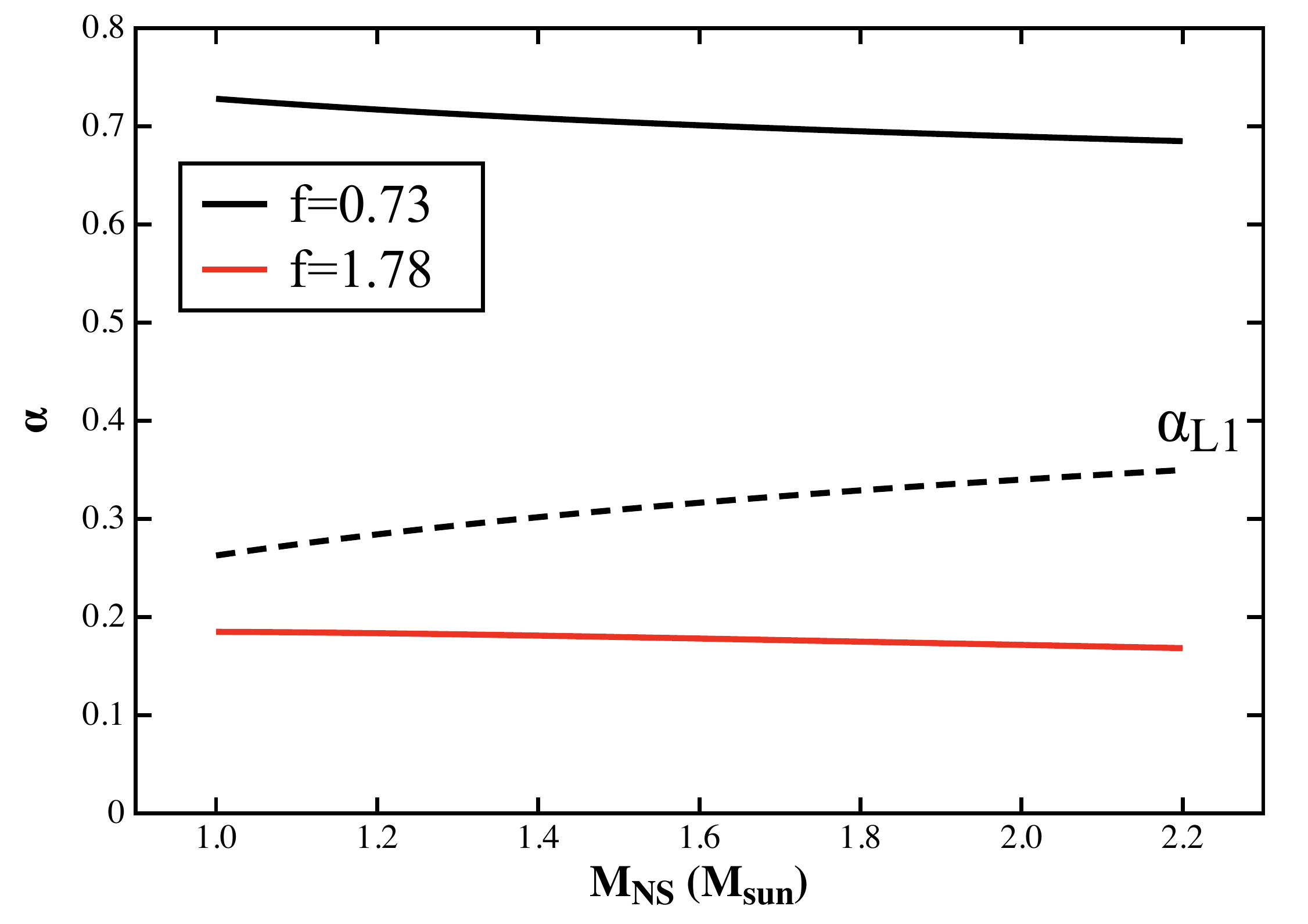}
\caption{Black and red solid lines represent the specific angular momentum as a function of the NS mass of the matter ejected by the system during the highly non-conservative mass transfer scenario calculated assuming the magnetic braking prescription of \citet[][f=0.73]{Skumanich:1972vy} and \citet[][f=1.78]{Smith:1979vn}, respectively. The parameter $\alpha$ is reported in units of the specific angular momentum of the donor. The dashed line shows the specific angular momentum calculated at the inner Lagrangian point L1 of the binary system.}
\label{fig:alpha}
\end{figure}

Following \citet{Burderi:2010tk}, the angular momentum loss due to GR and MB can be summarised in the expression:
\begin{equation} 
\label{eq:GRMB}
\left(\frac{\dot{J}}{J_{\rm orb}}\right)_{\rm GR+MB}=-11.4\times 10^{-19}m_1~m_{2}~m^{-1/3}P_{\rm orb,9h}^{-8/3}[1+T_{\rm MB}],
\end{equation}
where $T_{\rm MB}$ is a parametrisation of the torque associated with the magnetic braking that plays a pivotal role in driving the mass transfer in binary systems characterised by orbital periods larger than a few hours \citep{Verbunt:1993vj}. In line with \citet[][]{Iaria:2018tq}, the parameter can be written as:
\begin{equation}
\label{eq:TMB}	
T_{\rm MB}=1.7\times 10^{-2}(k^2)_{0.1}f^{-2}m_1^{-1}P_{\rm orb,9h}^2q^{1/3}(1+q)^{2/3},
\end{equation}
where $k$ is the gyration radius of the secondary star and $f$ represents a dimensionless parameter for which we assumed two preferred values being $f=0.73$ \citep{Skumanich:1972vy} and $f=1.78$ \citep{Smith:1979vn}.
Combining Eq.~\ref{eq:m2_pdot},~\ref{eq:evomom},~\ref{eq:GRMB}, and ~\ref{eq:TMB}, we can determine the specific angular momentum $\alpha$ that matter ejected from the system at a rate $-(1-\beta)\dot{M}_2$ should carry in order to generate an expansion of the orbital period at the observed $\dot{P}_{\rm orb}$ for \swiftj{}. In Figure~\ref{fig:alpha} we report the parameter $\alpha$ as a function of the NS mass in the range 1.0--2.2 M$_\odot$. We note that the secondary mass values required in the relations described above have been inferred following \citet{Markwardt:2010tl}. Solid lines in Figure~\ref{fig:alpha} have been determined taking into account different prescriptions for the angular momentum losses via magnetic braking, more specifically the black line assumes $f=0.73$ \citep{Skumanich:1972vy} and the red lines $f=1.78$ \citep{Smith:1979vn}, while for both we fixed $k=0.35$ that represents the gyration radius for a 0.7~M$_\odot$ donor star \citep[see e.g.,][]{Claret:1990to}. As a reference point, we also reported the specific angular momentum of matter sitting at the inner Lagrangian point L1 (dashed line) and calculated it as $\alpha_{L1}=[1 -0.462(1 + q)^{2/3}q^{1/3}]^2$. It is noteworthy that both solutions prescribe matter being ejected in between the primary and secondary stars of the system, however, the prescription of \citet{Smith:1979vn} implies mass leaving the system inside the Roche-lobe of the primary while Skumanich's prescription suggests the ejection close to the donor star.

A possible mechanism proposed to explain highly non-conservative mass transfer in binary systems harbouring a fast-rotating NS requires the crucial role of the radiation pressure of the magneto-dipole rotator generated by the NS in quiescence \citep[see e.g.,][]{Burderi:2003wk,di-Salvo:2008uu}. Indeed, radiation pressure generated by the magneto-dipole may be able to push away matter transferred by the donor around the inner Lagrangian point \citep[radio-ejection, see e.g.,][for a detailed discussion on the model]{Burderi:2001wp, di-Salvo:2008uu}. Any time the pressure balance between radiation and overflowing matter from the donor breaks due to a significant increase of the latter, X-ray outbursts might set in. This mechanism has been suggested to explain fast orbital expansions in the AMXPs SAX J1808.4$-$3658 \citep[see e.g.,][]{di-Salvo:2008uu,Burderi:2009td,Sanna:2017vj,Tailo:2018aa} and SAX J1748.9$-$2021 \citep[][]{Sanna:2016ty}. Similar conclusions have been drawn for the eclipsing binary systems X1822$-$371 \citep{Burderi:2010tk,Iaria:2015ut} and MXB 1659-298 \citep{Iaria:2018tq}, both showing large secular orbital period derivatives. Finally, it should be emphasised that this scenario might not be suitable for the specific case of \swiftj{}. Firstly, Figure~\ref{fig:alpha} suggests that matter is ejected from the system far from the inner Lagrangian point, however, detailed calculations are required to verify possible equilibrium in such different conditions. Secondly, in order to match the predicted donor mass-transfer rate required to explain the orbital expansion at the bolometric value extrapolated from the X-ray outburst, we must invoke a mechanism that allows a very small fraction of transferred mass ($\sim0.2-0.3\%$) to reach the NS during the outburst phase (possible dimming effects due to the high inclination angle of the source should be taken into account for a more accurate estimation of this parameter). On the other hand, the radio-ejection model is usually interpreted as an on/off mechanism that allows or prevents the whole mass-transfer rate to be accreted. Furthermore, we can roughly estimate the strength of the NS dipolar magnetic field required to stop the predicted mass-transfer rate at the inner Lagrangian point L1. Following \citet{Burderi:2002wk}, we can equate the ram pressure of the overflowing matter $P_{\rm RAM}=1/2 \rho v_{\rm ff}$ (where $\rho = 4 \pi R^2 v_{\rm ff}=\dot{M}_2$ is the density of the overflowing matter, and $v_{\rm ff}=(2 G M_1/R)^{1/2}$ is the free-fall velocity at the distance $R$), and the radiation pressure of the rotating magnetic dipole, assumed isotropic, $P_{\rm PSR}=2.04\times 10^{12} P^{-4}_{\rm ms}\mu^2_{26}R^{-2}_{6}\,\text{dy}\,\text{cm}^{-2}$ (where $P_{\rm ms}$ is the NS spin period in milliseconds, $\mu_{26}$ is the magnetic moment in units of $10^{26}\text{G}\,\text{cm}^3$, and $R_{6}$ is the distance from the NS surface in units of $10^{6}$~cm). We can then derive the expression of the magnetic dipole as:
\begin{equation}           
\mu_{26}\simeq20 R^{-1/2}_{10} m_1^{1/2}P^4_{\rm ms}\dot{m}_{2,-7} ~\text{G}~\text{cm}^3,
\end{equation}
where $R_{10}$ represents the distance in units of $10^{10}$~cm, and $\dot{m}_{2,-7}$ is the secondary mass-transfer rate in units of $10^{-7}\,{M}_\odot\,\text{yr}^{-1}$. Considering the value of $\dot{m}_2$ estimated from the fast orbital expansion of the binary system, we obtain $\mu_{26}\sim 20~\text{G}~\text{cm}^3$, which implies a dipolar magnetic field $B\sim 1.3\times 10^9$~G, almost an order of magnitude larger than the values inferred from the long-term spin evolution of the system reported in Sec.~\ref{sec:spinevo}.

\paragraph{Gravitation quadrupole coupling} 
\label{sec:gqc}
Orbital period modulations have been observed in several millisecond radio pulsars \citep[e.g., PSR B1957$+$20, PSR J2051$-$0827, and PSR J2339$-$0533;][]{ Applegate:1994vp,Doroshenko2001a,Pletsch2015a}, and interpreted as due to gravitational quadrupole coupling (GQC) between the companion star and the orbit \citep{Applegate:1992uh, Applegate:1994vp}. The model invokes changes in the mass distribution of the donor star triggered by magnetically induced transitions between different states of fluid hydrostatic equilibrium. Changes of the donor oblateness translate into changes in the gravitational force between the binary components inducing orbital period variations on dynamical timescales. As reported by \citet{Applegate:1992uh}, it is possible to express the variation of the orbital period as a function of the variable quadrupole moment $\Delta Q$ as:
\begin{eqnarray}
\frac{\Delta P_{\rm orb}}{P_{\rm orb}}= -9\frac{\Delta Q}{M_2 a^2},
\label{eq:ap1}
\end{eqnarray}
where $M_2$ is the companion mass and $a$ is the orbital separation. 
$\Delta Q$ represents the variable part of the donor quadrupole moment that depends on the variation of its outer layers angular velocity $\Delta \Omega$ due to angular momentum transfer $\Delta J$ such that $\Delta \Omega=\Delta J /I_s$, where $I_s$ represent the moment of inertia of the outer layers. 
Under the assumption that the star angular velocity $\Omega$ is locked to the orbital angular velocity \citep{Applegate:1994vp}, we can estimate the required variable angular velocity to induce orbital period changes $\Delta P_{\rm orb}$:
\begin{eqnarray}
\frac{\Delta \Omega}{\Omega}= \frac{G\,M_2^2}{2R_2^3\,M_s} \left(\frac{a}{R_2}\right)^2\left(\frac{P_{\rm orb}}{2\pi}\right)^2 \frac{\Delta P_{\rm orb}}{P_{\rm orb}},
\label{eq:ap3}
\end{eqnarray}
where $M_s$ is the mass of the secondary thin shell rotating with angular velocity $\Omega$. Following \citet{Applegate:1992uh}, the variable part of the donor luminosity $\Delta L$ required to generate changes in the orbital period $\Delta P_{\rm orb}$ can be expressed as:
\begin{eqnarray}
\Delta L\simeq \frac{\pi}{3} \frac{G\,M_2^2}{R_2 P_{\rm mod}} \left(\frac{a}{R_2}\right)^2 \frac{\Delta \Omega}{\Omega} \frac{\Delta P_{\rm orb}}{P_{\rm orb}},
\label{eq:ap5}
\end{eqnarray}
where P$_{\rm mod}$ represents the period of the orbital modulation. 
The application of this scenario for \swiftj{} is potentially limited by the fact we currently do not detect a long-term orbital modulation due to the very limited number (only two at the moment) of observed outbursts from which the NS ephemeris can be determined. Notwithstanding, we can estimate, at least at zero-order, the energetic reservoir required for the source. To do that, we speculate that the orbital period variation $\Delta P_{\rm orb}\sim0.140$~s occurred over a time span of almost eleven years represents the first quarter of a sinusoidal modulation characterised by a $P_\text{mod}\sim 44$~yr and amplitude equal to $\Delta P_{\rm orb}$. Furthermore, we assume that the outer layer that differentially rotates with respect to the donor star has a mass $M_s\simeq 0.1 M_2$ \citep{Applegate:1994vp}, the donor  Roche-Lobe radius $R_{L2}\simeq 0.462\, [q/(1+q)]^{1/3} a$ \citep[valid for mass ratio $q\leq0.8$;][]{Paczynski71} is a good proxy of $R_2$	and the internal reservoir of energy available to support the outer layer oscillations must be of the order of 10\% of the thermonuclear energy produced by the donor \citep{Applegate:1992uh}. Using the aforementioned assumptions in Eq.~\ref{eq:ap3} and Eq.~\ref{eq:ap5}, and rescaling the parameters by the properties of \swiftj{}, we find that the donor star must have a luminosity $L\sim0.36L_\odot$ to justify the observed orbital period variation employing the GQC mechanism. Considering a donor star mass $M_2\sim0.7$~M$_\odot$, we can assume $L_2/L_\odot=(M_2/M_\odot)^{4.5}$ that implies $L_2\simeq0.2$~L$_\odot$. We find a discrepancy close to a factor of two between the required and available energy values, however, it should be stressed that some of the assumptions discussed earlier for \swiftj{} remain speculations (e.g., we stress that the modulation period suggested could be either significantly shorter or longer) at the moment of writing this work. Further observations of the source in outburst are therefore needed to additionally investigate this scenario.
Finally, under the assumption that the exchange of angular momentum between the core and outer layers of the donor is driven by a magnetic torque exerted by a subsurface field, we can estimate its value to be close to $10^4$~G \citep[see][equation 23]{Applegate:1992uh}.

\subsubsection{Projected semi-major axis}
\label{sec:asin}
According to Kepler's third law, we expect changes in the orbital period of a binary system to be accompanied by changes of the orbital separation $a$ and/or on the total mass of the system. As described in Sec.~\ref{sec:orb}, \swiftj{} showed a fast orbital expansion during the last decade. Moreover, from Table~\ref{tab:solution} we can deduce an increase of the NS projected semi-major axis between the observed outbursts of $\Delta x=8.8(2.3)\times 10^{-5}$~lt-s, corresponding to an expansion rate of $\dot{x}=2.56(32)\times 10^{-13}\,\text{lt-s}~\text{s}^{-1}$. Even if the inferred rate results are marginally significant, it should be emphasised that this is the first time we directly measure a clear evolution of the NS projected semi-major axis for AMXPs. A comparison between the evolution timescales shows that $x$ expands almost a factor of ten faster ($x/\dot{x}\sim 0.2$~Myr) with respect to the $P_{\rm orb}$ ($P_{\rm orb}/\dot{P}_{\rm orb}\sim 2.4$~Myr). 
Given the observed values of $\dot{x}$ and $\dot{P}_{\rm orb}$ we can predict the mass transfer rate required to verify instantaneously Kepler's third law. To do that, we calculate Kepler's third law logarithmic derivative as:
\begin{equation}
\label{eq:kep}
3\frac{\dot{a}}{a}-2\frac{\dot{P}_{orb}}{P_{orb}}=\frac{\dot{M}_1+\dot{M}_2}{M_1+M_2}.
\end{equation}  
where $\dot{a}$ represents the time derivative of the orbital separation. Since $x=a_1 \sin(i)/c$ defines the projected semi-major axes of the NS orbit expressed in light-seconds, and considering that $a_1=aM_2/(M_1+M_2)$, we can re-write Eq.~\ref{eq:kep} as follows:
\begin{equation}
\label{eq:kep2}
3\frac{\dot{x}}{x}-2\frac{\dot{P}_{\rm orb}}{P_{\rm orb}}=3\frac{\dot{M}_2}{M_2}-2\frac{\dot{M}_1+\dot{M}_2}{M_1+M_2}.
\end{equation}
If we assume a conservative mass transfer, $\dot{M_2}=-\dot{M}_1$, the total mass of the system must remain constant. If that is the case, the second term on the right-hand side of Eq.\ref{eq:kep2} is zero. Moreover, since $\dot{x}/{x} \sim 10\,\dot{P}_{\rm orb}/P_{\rm orb}$, we find that $\dot{x}/{x}\simeq \dot{M}_2/{M_2}$. Therefore, if the observed fast expansion of the NS orbit is related to the exchange of matter within the binary system, the donor mass transfer rate should be of the order of $4.2\times 10^{-6}~\text{M}_{\odot}~\text{yr}^{-1}$ for a $0.7~\text{M}_{\odot}$ companion star. The predicted mass transfer rate is more than an order of magnitude larger than the value predicted from the fast expansion of the observed orbital period discussed in Sec.~\ref{sec:orb}. Two considerations should be discussed with respect to this result: firstly, the very short timescale $M_2/\dot{M}_2\sim0.2\,\text{Myr}$ would make it very unlikely to catch the system in action; secondly, the fact that during the outburst phase we infer a mass-transfer rate orders of magnitude smaller makes the conservative mass-transfer hypothesis inapplicable. However, even in the hypothesis of a highly non-conservative mass-transfer scenario (with $\beta$ of the order of a few \%), it can be easily shown that Eq.~\ref{eq:kep2} still implies $\dot{x}/{x}\simeq \dot{M}_2/{M_2}$ if we consider the possible range of masses of the NS and companion for \swiftj{}. 

It could be interesting to at least mention a few alternative scenarios that could in principle be invoked to explain secular changes of orbital parameters. The first mechanism, discussed by \citet{Kopeikin:1996wy}, takes into account the proper motion of a binary pulsar and its effect on changing the observed geometrical orientation of the orbital plane. Since the effect is purely geometrical, no real variation of the binary orbit is expected, therefore secular changes are expected for $x$, not for the orbital period. Following \citet{Arzoumanian:1996up}, we can quantify the effect with the relation $\dot{x}/x=\mu \sin{(j)}/\tan{(i)}$, where $\mu$ represents the proper motion magnitude, $i$ is the orbital inclination and $j$ is the angle defined by the line of nodes and the proper motion vector. For $i=77^{\circ}$, we find that $\mu \sin j\sim 3\times 10^{-13} \text{rad}\, \text{s}^{-1}\simeq 1950\,\text{mas}\, \text{yr}^{-1}$, which corresponds to a proper speed of $\approx 65000$ km s$^{-1}$! for an object almost $7$~kpc apart and located in the Galactic Plane. 

The second, more attractive, scenario that is worth mentioning implies the presence of another object bound to the binary system, forming a hierarchical triple system. If correct, $\dot{x}$ could arise from the precession of the binary system in the gravitational field of the third body. In this scenario, it is reasonable to expect effects also in the temporal evolution of the orbital period that will translate into $\dot{P}_{orb}$. A few multiple systems harbouring millisecond radio pulsars have been discovered in the past such as B1620-26, PSR B1757+12, and PSR1257 + 12 \citep[see e.g.,][]{Thorsett:1999vo, Wolszczan:1992wd, Ransom:2014vb}. To get a sense of the possible effects on the secular variation of the binary orbital parameters influenced by a lighter body orbiting around, it is interesting to mention the system B1620-26. This system, most likely composed of a 1.3~M$_\odot$ NS and a 0.3~M$_\odot$ white dwarf, is surrounded by a 0.01~M$_\odot$ star locked in a wide orbit with an orbital period of order one hundred years. Interestingly, by investigating the secular temporal properties of the pulsar on timescales shorter than the outer body orbital period, the system shows a variation of the orbital parameter such as $\dot{x}\sim -6\times 10^{-13}\,\text{lt-s}~\text{s}^{-1}$ and $\dot{P}_{orb}\sim4\times 10^{-10}\,\text{s}~\text{s}^{-1}$ \citep{Thorsett:1999vo}, similar in order of magnitude with those discussed here for \swiftj{}. The available knowledge on the long-term orbital behaviour of the source presented here allows us to only speculate on these possible alternative scenarios, however, new X-ray outbursts will allow us to set tighter orbital constraints useful to improve the investigation.

\subsection{Spin evolution}
\label{sec:spinevo}
To investigate possible spin-down down effects during the quiescence phase of \swiftj{}, we compared the spin frequency at the end of the 2010 outburst with the value at the beginning of the latest one. As mentioned in Sec.~\ref{sec:results}, to consistently compare the NS spin evolution between the two outbursts, we re-analysed the \rxte{} dataset presented by \citet{Markwardt:2010tl} and \citet{Altamirano:2011uq} to take into account a more accurate set of coordinates of the source obtained with \chandra{} during its quiescence state \citep{Jonker:2013wp}. As previously discussed, positional uncertainties play an important role in accurately determining the spin frequency as well as its temporal derivative. 
If we take into account the upper limit on the spin frequency derivative during the 2010 outburst and we propagate the uncertainties on the spin frequencies of the two outbursts, we infer an upper limit for the spin variation during quiescence of $|\Delta \nu|\lesssim7.2\times 10^{-7}~\text{Hz}$ (95\% c.l.) and therefore a spin frequency derivative $|\dot{\nu}|\lesssim 2.1\times 10^{-15}~\text{Hz}~\text{s}^{-1}$ (95\% c.l.).
Adopting a description of the pulsar magnetosphere based on the force-free relativistic MHD solution proposed by \citet{Spitkovsky:2006uz}, we can constrain the NS magnetic dipole moment to be:
\begin{equation}
\mu < 0.7\times10^{26}\left(\frac{1}{1+\sin^2{\alpha}}\right)^{1/2} I_{45}^{1/2}\nu_{518}^{-3/2}\dot{\nu}_{-15}^{1/2}\,\,\,\, \text{G}~{cm}^3,
\label{eq:mag}
\end{equation}
where $\alpha$ is the angle between the rotation and magnetic poles, $I_{45}$ is the NS moment of inertia in units of $10^{45}$~g~cm$^2$, $\nu_{518}$ is the NS spin frequency rescaled to match \swiftj{}, and $\dot{\nu}_{-15}$ is the spin-down frequency derivative in units of $10^{-15}\,\text{Hz}\,\text{s}^{-1}$. We can then calculate an upper limit on the NS magnetic moment by assuming the extreme value $\alpha=0$~deg, which corresponds to $\mu < 1\times10^{26}\,\,\, \textrm{G\,cm$^3$}$.
Defining the magnetic field strength at the magnetic caps as $B_{PC}= 2 \mu/R_{NS}^{3}$, and considering a NS radius of $R_{NS}=1.14\times10^{6}$ cm  \citep[corresponding to the FPS equation of state for a 1.4~M$_\odot{}$ NS, see e.g.,][]{Friedman1981a, Pandharipande1989a}, we obtain $B_{PC}<1.3\times 10^{8}\,\text{G}$, in line with the estimates reported by \citet{Mukherjee:2015td} and similar to what has been derived for other AMXPs. Future outbursts are required to improve the constraint on the magnetic field of the NS.

\section{Conclusions}
We investigated the temporal properties of the coherent X-ray pulsation from \swiftj{} during its latest outburst using data collected by \nicer{} and \xmm{}, and we compared it with the \rxte{} dataset of the previous outburst observed almost eleven years earlier that we re-analysed. We found the following:

\begin{enumerate}[i)]
  \item The pulse profile is well described by the superposition of two harmonically related sinusoidal components with a comparable fractional amplitude which varies during the outburst. Large phase jumps are shown by the fundamental phase of the signal (Sec.~\ref{sec:phase}) consistently to what was observed during the previous outburst. On the other hand, only a small phase wandering consistently with typical timing noise observed in AMXPs is observed for the second harmonic. Interestingly, phase jumps seem to be accompanied by an increase of harmonic content in the pulse profile. 
  \item Both ephemerides obtained from the timing analysis of the detected outbursts show some marginally significant evidence for non-zero eccentricity ($e\simeq4\times 10^{-5}$, Sec.~\ref{sec:ecc}), which currently makes \swiftj{} the only AMXP with a constraint on this parameter.   
  \item Long-term evolution of the orbital parameters of \swiftj{} suggests a fast expansion of the projected semi-major axis ($x$), as well as the orbital period ($P_{\rm orb}$), at a rate of $\dot{x}\simeq 2.6\times 10^{-13}\,\text{lt-s}\,\text{s}^{-1}$ (Sec.~\ref{sec:asin}) and $\dot{P}_{\rm orb}\simeq 4 \times 10^{-10}\,\text{s}\,\text{s}^{-1}$ (Sec.~\ref{sec:orb}), respectively. If interpreted in terms of loss of mass and orbital momentum in the binary system, these results would imply a highly non-conservative mass-transfer process in which the donor star transfers matter at a super-Eddington rate and only a tiny fraction of the transferred matter ($< 0.5\%$) accretes onto the compact object (Sec.~\ref{sec:ncmt}), which makes this scenario unlikely for the specific case of \swiftj{}. On the other hand, the fast orbital expansion could be explained, at zero-order, as the result of a gravitational quadrupole coupling between the companion star and the orbit (Sec.~\ref{sec:gqc}). Finally, the speculative scenario invoking a hierarchical triple system cannot be ruled out at the moment.  
  \item Secular evolution of the spin frequency does not show a significant deceleration of the compact object. The upper limit on the spin derivative allows us to constrain the magnetic field strength at the polar caps as $B_{PC}<1.3\times 10^{8}\,\text{G}$ (Sec.~\ref{sec:spinevo}), which is in line with typical values reported for AMXPs.  
\end{enumerate}

\section*{Acknowledgements}
PB acknowledges support from the CRESST II cooperative agreement (80GSFC21M0002). AM acknowledges a financial contribution from the agreement ASI-INAF n.2017-14-H.0 and from the INAF mainstream grant (PI: T. Belloni, A. De Rosa). AM, TDS, AA and RI acknowledge financial contribution from the HERMES project financed by the Italian Space Agency (ASI) Agreement n. 2016/13 U.O. AM is supported by the H2020 ERC Consolidator Grant “MAGNESIA” under grant agreement No. 817661 (PI: Rea) and National Spanish grant PGC2018-095512-BI00. This work was also partially supported by the program Unidad de Excelencia Maria de Maeztu CEX2020-001058-M, and by the PHAROS COST Action (No. CA16214).

\section*{Data availability }
The data utilized in this article are publicly available at \href{https://heasarc.gsfc.nasa.gov/cgi-bin/W3Browse/w3browse.pl}{https://heasarc.gsfc.nasa.gov/cgi-bin/W3Browse/w3browse.pl}, while the analysis products will be shared on reasonable request to the corresponding author.


\bibliographystyle{mnras}
\bibliography{biblio.bib}







\bsp	
\label{lastpage}
\end{document}